\documentclass[aps,pra,twocolumn]{revtex4-1}
\usepackage{epsfig,amssymb,amsmath,mathrsfs,amsthm,graphicx,float,color}
\usepackage[active]{srcltx}
\usepackage{subfigure}
 \def\map#1{\mathcal #1}
\def\<{\langle}\def\>{\rangle}

\def\Tr{\operatorname{Tr}}\def\:{\hbox{\bf
    :}}

\def\grp#1{\mathsf{#1}}
\def\st#1{\mathbf{#1}}

\def\spc#1{\mathscr{#1}}

\newtheorem{theo}{{Theorem}}

\newtheorem{lem}{{Lemma}}

\begin{document}
\title{Universal super-replication   of unitary  gates}
\author{G. Chiribella, Y. Yang and C. Huang} 
\affiliation{Center for Quantum Information, Institute for Interdisciplinary Information Sciences, Tsinghua University, Beijing 100084, China   }
\begin{abstract}
Quantum states    obey  an asymptotic no-cloning theorem, stating that  no deterministic machine can reliably replicate  generic sequences of identically prepared pure states.  
   In stark contrast, we show that generic sequences   of unitary gates  can be  replicated deterministically at nearly quadratic rates,   with   an error  vanishing on most inputs except for an exponentially small fraction.
    The result is not in contradiction with the no-cloning theorem, since the impossibility of  deterministically transforming  pure states into unitary gates prevents  the application  of the gate replication  protocol to  states. 
In addition to gate replication,    we show  that  $N$ parallel uses of a completely unknown unitary  gate can be compressed  into a single gate acting on $O(\log N)$ qubits, leading to  an exponential reduction of the amount of quantum communication needed to implement the gate  remotely.   \end{abstract}
\maketitle

A striking feature of quantum theory is the impossibility of constructing a universal copy machine, which takes as input a quantum system in an arbitrary   pure state  and produces as output a number of exact replicas   \cite{wootters,dieks}.    
Such an impossibility   has  major implications for quantum error correction   \cite{gottesman2009introduction,lidar2013errorcorrection} and cryptographic protocols such as   key distribution \cite{cloning-RMP,review-cerf}, quantum secret sharing \cite{hillery,gottesmann-lo}, and quantum money \cite{quantum-money,Aaronson09,FarhiGosset12,MolinaVidick13}.   

The impossibility of universal copy machines   is a hard fact: it equally affects   deterministic  \cite{buzek-hillery,gisin-massar,werner,keyl-werner}    and probabilistic  machines  \cite{duan-guo,fiurasek,stateREP},  whose performances coincide with those of deterministic  machines  when it comes to copying  completely unknown  pure states     \cite{fiurasek,stateREP}.      
  A similar no-go result holds   when the universal machine is presented a large number  $N$ of identical  copies    and is required to produce a larger number  $M>N$ of  approximate replicas:  if the replicas have  non-vanishing overlap  with the desired $M$-copy state, then the number of extra copies   must be negligible compared to $N$ \cite{werner}.  We refer to this fact as the \emph{asymptotic no-cloning theorem}, expressing the fact that independent and identically distributed (i.i.d.) sequences of pure states cannot be stretched by any significant amount.   
  The asymptotic no-cloning theorem holds 
   also for non-universal machines designed to copy  continuous sets of states, 
  provided that such machines work deterministically \cite{stateREP}. 
  
The impossibility of universal state cloning  suggests similar results for quantum gates.      Along this line, a no-go theorem for universal gate cloning was proven in Ref. \cite{gate-no-cloning},   showing that no quantum network can perfectly simulate two uses of an unknown unitary gate by querying it  only once.   Optimal  networks that approximate universal gate cloning were studied in Refs. \cite{gate-no-cloning,bisio2014optimal}.   Very recently, D\"ur and coauthors \cite{duer} considered a  non-universal setup designed to  clone  phase gates,~i.~e.~gates generated by time evolution with a known Hamiltonian.   In this scenario, they  devised  a quantum network that approximately simulates  up to $N^2$ uses of an unknown phase  gate while using it  only $N$ times, with vanishing  error  in the  large $N$ limit.  
        Such a result  establishes the possibility  of \emph{super-replication} of phase gates---super-replication being   the generation of $M\gg N$ high-fidelity replicas from $N\gg 1$ input copies \cite{stateREP}.   
   Remarkably, super-replication of phase gates is achieved deterministically,  whereas super-replication of phase states has  exponentially small probability of success.
    The main  open question raised by Ref. \cite{duer}  is whether deterministic super-replication    occurs not only for phase gates, but also for  arbitrary  unitary gates.  An affirmative answer would imply that the asymptotic no-cloning theorem only applies to states, whereas it  is possible to stretch long i.i.d. sequences of reversible  gates by up to a quadratic  factor.  

In this letter we answer the question in the affirmative, establishing the possibility of universal super-replication of unitary gates, in stark contrast with the asymptotic no-cloning theorem for pure states.     Given $N$ uses of a completely unknown unitary gate $U$, we construct a quantum network that simulates up to $N^2$ parallel uses of $U$, providing an output that is close to the ideal target for all possible input states except for an exponentially small fraction. 
 The quadratic replication rate is optimal: every  other network   producing replicas  at a  rate higher than quadratic will necessarily  spoil their quality,  delivering an output   that has vanishing overlap with the output of the desired gate. 
    In addition to replication, we consider the  task of  \emph{gate compression}, where the goal is  to faithfully encode the action of a black box into a gate operating on a smaller quantum system.      
 We show that $N$ uses of a completely unknown gate can be encoded without any loss into a single gate acting only on $   (d-1)  (d/2 +1)\log  N$ qubits, thus allowing for an exponential reduction of computational workspace.  The number of qubits can be further cut down by a half  if one  tolerates an error that vanishes on almost all inputs in the large $N$ limit.
  The compression  of i.i.d. gate sequences is the analogue of the  compression of  i.i.d. state sequences  \cite{StateCompression}, recently  demonstrated  experimentally \cite{Compression}.

{\em Universal super-replication of qubit gates.}   Let us start from the simple case of qubit gates,   represented by unitary matrices  in $\grp{SU}  (2)$. 
A generic gate can be parametrized as $U_{\theta,\st n}   =  \exp[  -  i  \theta   \,  \st n\cdot  \st j]$, where $ \theta  \in  [0,2\pi)$ is a rotation angle, $\st n=   ( n_x,n_y,n_z)$ is a rotation axis, and $\st j  =  (j_x,j_y,j_z)$ is the vector of angular momentum operators  ($j_i  :=  \sigma_i/2 \,  ,  i  =  x,y,z$).   We  define $g:  =  (\theta,  \st n)$ and label the unitary gate as $U_g$.  
Here   both $\theta$ and $\st n$ are completely unknown, differing from the setting of \cite{duer}, where the rotation axis was fixed and only $\theta$ was varying.

\begin{figure}
  \centering
  \includegraphics[width=1.02\linewidth]{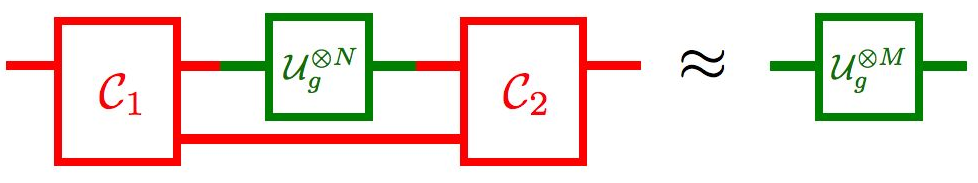}\\
  \caption{{\bf Quantum network for gate replication.}  The network simulates $M$ parallel uses of an unknown unitary  gate $\map U_g$, while querying it  only $N$ times.  The simulation is obtained by transforming 
     the input state of $M$ systems into the joint  state of $N$ systems plus an ancilla (via quantum channel $\map C_1$), applying the unknown gate on the $N$ systems, and then recombining them with the ancilla via  a quantum channel $\map C_2$, which finally produces $M$ output systems.  }  \label{fig:gatecloning}
  \end{figure}

 In order to replicate  unknown gates, we consider a network where $N$ parallel uses     of   $U_g$   are sandwiched between two quantum channels, $\map C_1$ and $\map C_2$, as in figure \ref{fig:gatecloning}.      The overall action of the network is described by the channel $   \map C_2  \, \left(  \map U_g^{\otimes N}  \otimes \map I_{A}  \right) \, \map C_1$, with $\map U_g(\cdot )   =  U_g\cdot   U_g^\dag$ and $\map I_A$ denoting the identity  on a suitable ancillary system.     
 To construct the channels $\map C_1$ and $\map C_2$,  we decompose the Hilbert space of $K$   qubits  ($K  =  N, M$) into  rotationally invariant subspaces.  
Choosing $K$ to be even, we have 
\begin{align}\label{spacedecomp}
\spc H^{\otimes K}   \simeq \bigoplus_{j=0}^{K/2}    \,  \left(    \spc R_j  \otimes \spc M_{jK}  \right)  \, ,
\end{align}
where $j$ is the quantum number of the total angular momentum,   $\spc R_j$ is a representation space,
of dimension $d_j=  2j+1$, and $\spc M_{jK}$ is a multiplicity subspace,  of dimension  $m_{jK}$.
The isomorphism in Eq. (\ref{spacedecomp}) is called the \emph{Schur transform} and can be implemented efficiently in a quantum circuit \cite{SchurTransform,harrowthesis}.  We now  introduce a cutoff on the quantum number $j$ and define the  subspace  
\begin{align}
\spc H^{(K)}_J     =   \bigoplus_{j\le J}   \left( \spc R_j  \otimes \spc M_{jK}\right)   \, .
\end{align}   
To compress a state inside this subspace, we  use the  encoding channel defined by   
\begin{align}
\map E^{(K)}_J(\rho)  :=    P^{(K)}_J   \rho  P^{(K)}_J     +    \Tr  \left[ \left(  I^{\otimes K}   -  P^{(K)}_J   \right)  \rho \right]  \,  \rho_0  \,   
\end{align} 
where $P^{(K)}_{J}$  is the projector on $\spc H^{(K)}_J$ and $\rho_0$ is a fixed density matrix with support in $\spc H^{(K)}_J$.
The key observation is that most $K$-qubit states are left nearly unchanged by the channel $\map E^{(K)}_J$, provided that $K$ is large and $J$ is large compared to $\sqrt K$.    
Denoting  by  $  F_\Psi^{(K,J)}$  
   the fidelity  between a generic  $K$-partite  pure state  $|\Psi\> $  and its compressed version $  \map E^{(K)}_J  (  |\Psi\>\<\Psi|)$, we have the following 
\begin{theo}\label{theo}   
 If $|\Psi\>$ is chosen uniformly at random, then, for every fixed $\epsilon>0$,  the probability that   $F_\Psi^{(K,J)}$   is smaller than $1-\epsilon$ satisfies the bound 
\begin{align}\label{smallprob}
{\sf Prob}   \left[ F^{(K,J)}_\Psi   <  1-\epsilon \right]  <    \frac{  
2 (K+1)  }{\epsilon }  \, \exp \left[-\frac{2J^2}{K}\right] \, . 
\end{align}
\end{theo}

{\bf Proof.} By Markov's inequality, one has  ${\sf Prob}   \left[ F^{(K,J)}_\Psi   <  1-\epsilon \right]    <   \,   (1-   \mathbb E\left[  F^{(K,J)}  \right])/\epsilon$, where  $\mathbb E\left[  F^{(K,J)}\right] $ is the average of the fidelity over all pure states.   In turn, the average fidelity  can be  lower bounded  by the entanglement fidelity    \cite{Horodecki,Nielsen}, given by  $  F^{(K,J)}_E  =   \<  \Phi_{2^K}| \, \left(   \map E^{(K)}_J  \otimes \map I^{\otimes K}  \right)   (|\Phi_{2^K}\>\<\Phi_{2^K}|)   \,  |\Phi_{2^K}\> $,  
 where   $|\Phi_{2^K}\>$ is a maximally entangled state in $\mathbb C^{2^K}\otimes \mathbb C^{2^K}$.      The entanglement fidelity satisfies  the bound  
\begin{align*}
\nonumber F^{(K,J)}_E  &\ge   \left |  \<   \Phi_{2^K}  |   \left( P_J^{(K)}\otimes I^{\otimes K}\right)  |\Phi_{2^K}\>  \right|^2 = \left[ \sum_{j\le J}   \frac{d_j m_{jK}}{2^K}\right]^2     \end{align*}
where the coefficients $d_j m_{jK}/2^K $ form a probability distribution,    known as the \emph{Schur-Weyl measure} \cite{meliot,biane}.  For large $K$, the Schur-Weyl measure concentrates   around $j=0$ \cite{harrowthesis}, yielding   the bound  $F^{(K,J)}_E  \ge 1-  2 (K+1)   \, \exp\left[-\frac{2J^2}{K}\right] $  (Appendix A), which combined with the previous observations implies  Eq. (\ref{smallprob}). \qed 

\medskip

Let us apply Theorem \ref{theo} to gate replication.  The theorem   guarantees that,  except for an exponentially small fraction, almost     all  pure $M$-qubit  states are approximately in a subspace  $\spc H_J^{(M)} $ with $J  \gg \sqrt M$.  To achieve gate replication,  we combine this fact with the observation that  for $J\le N/2$, the states  in $\spc H^{(M)}_J$ can be faithfully encoded into   $\spc H^{\otimes N}\otimes  \spc H_A$, where $\spc H_A$ the Hilbert space of a suitable ancilla. The encoding is achieved by an isometry     $V_J$  
that commutes with all rotations, namely   
 \begin{align}\label{comm}  V_J  \, U_g^{\otimes  M}  =    \left(U_g^{\otimes N}  \otimes I_A\right) \, V_J   \qquad \forall\ U_g \in  \grp{SU}  (2) \, .
 \end{align}    
 
We are now ready to specify the channels $\map C_1$ and $\map C_2$ in the gate replication network of figure \ref{fig:gatecloning}.    For channel
$\map C_1$, we choose   $  \map C_1   =      \map V_{J}  \map E^{(M)}_{J}$,  where $\map V_J$ is the isometric channel $\map V_J(\cdot)  =  V_J \cdot V_J^\dag$ and $J$ is set to   $J    =   \left    \lceil  \sqrt  {N^{1- \alpha/4}} \right \rceil$ for $M$ growing like $N^{2-\alpha}$,  $\alpha>0$ and to  $J=  N/2$  for $M$ growing like $N^2$ or faster.  
 For channel   $\map C_2$ we choose the inverse of $\map V_j$, namely
   \begin{align}
  \map C_2(\rho)   =   V_{J}^\dag  \rho  V_{J}    +      \Tr\left[  \left( I^{\otimes N}  -   V_{J}  V_{J}^\dag  \right)  \rho  \right]  \,  \rho_0 \,.
  \end{align}     
The action of the network on a generic $M$-qubit  state  $U_g$ is then given by  
\begin{align}
\nonumber  \map C_2   (\map U_g^{\otimes N}  \otimes \map I_A)  \map C_1   (|\Psi\>\< \Psi| )  &  =    \map U_g^{\otimes M}   \map C_2  \map C_1  (  |\Psi\>\<  \Psi|)  \\
 & =   \map U_g^{\otimes M}     \map E_{J}^{(M)} (   |\Psi\>\<  \Psi|  ) \, ,  \label{aaa}
 \end{align}
the first equality coming from  Eq. (\ref{comm}) and the second from  the fact that $\map C_2$ is the inverse of $\map V_J$. 
Clearly, Eq. (\ref{aaa}) implies that the fidelity between the output state  and the ideal target $U_g^{\otimes M}  |\Psi\>$  is equal to $F^{(M,J)}_\Psi$  independently of $g$.  By  theorem \ref{theo}, the fidelity is  arbitrarily close to one  on most input states whenever  $M $ grows like $O(N^{2-\alpha})$. In this case,   the number of ancillary qubits used by the protocol  scales like  $M-N +  O\left(  N^{1-\alpha/2} \right)$ (Appendix B).  For $M$ growing faster than $N^2$, the fidelity tends to zero.

In summary, given $N$ uses of a completely unknown gate, our network  simulates   up to $ N^2$ uses with high fidelity on most  input states.       The simulation works  with probability exponentially close to 1, but can fail on some specific inputs:  notably, it fails for inputs of the i.i.d. form $  |\varphi\>^{\otimes M}$, for which the fidelity with the desired output state is \emph{zero}.  The fact that replication works well in the typical case  is analog to other  phenomena based on measure  concentration, such as   quantum equilibration  \cite{popescu-winter-short}  and entanglement typicality \cite{entangleAspects}.

{\em State cloning vs state generation.}   Deterministic gate replication is not in contradiction with  
 the  asymptotic no-cloning theorem. 
  Indeed, suppose that we wanted to use  gate replication  to clone a completely unknown pure state $|\psi\>  =  U_\psi  |0\>$ for some fixed state $|0\>$.     To this purpose, we would have to retrieve the gate $U_\psi$ from the input state $|\psi\>$---a task whose deterministic  execution is forbidden by  
    the no-programming theorem \cite{nielsen1997programmable}.  
  Since the state  $|\psi\>$ is arbitrary, a  symmetry argument precludes  the conversion $|\psi\> \to U_\psi$  even probabilistically   \cite{stateREP}.  

Though gate replication cannot be used for state cloning, it may still provide an advantage in the less demanding task of \emph{state generation}, which    consists in  producing $M$ copies of the state $|\psi\>$ from $N$ uses of the  gate  $U_\psi$.     The advantage does not show up in the universal case,    because universal gate replication  does not reproduce correctly the action of the gate  $U_\psi^{\otimes M}$   on the  i.i.d. input state  $|0\>^{\otimes M}$.    However, it  does show up  in non-universal cases: for example,  Ref. \cite{duer} demonstrated  that $N$ uses of a phase gate allow one to generate up to $N^2$ copies of the corresponding phase state.        Similarly, we show that universal gate replication allows to generate $M$   maximally entangled states with   fidelity 
\begin{align}
F^{\rm ent}_{\rm gen}  [N\to M]  \ge 1-  2  (M+1)\,  \exp\left[-\frac{N^2}{2M}\right]  
\end{align} 
(Appendix C). 
   The protocol for generating  entangled states also provides an alternative way to generate   phase states: given $N$ uses of a phase gate with phase $\theta$  one can first generate $M\gg N$ approximate copies of the maximally entangled state  $ ( |0\>|0\>   +  e^{-i \theta}|1\>  |1\>  )/\sqrt 2$ and then apply a CNOT gate to each copy  and discard the second qubit of the pair, thus obtaining $M\gg N $ approximate copies of the state $( |0\>  +  e^{-i\theta}  |1\>  )/\sqrt 2$.   We refer to the ability to produce $M\gg N$  states from $N\gg 1$ uses of the corresponding gate as  \emph{state super-generation}. Again, we stress that state super-generation does not challenge the asymptotic no-cloning theorem, because    $N$ uses of a gate  cannot be obtained deterministically from    $N$ copies of the corresponding state.  In the concrete examples of phase states and maximally entangled states,  the fidelity of the best  deterministic   $N$-to-$M$ cloner    scales    as  $(N/M)^{1/2}$ and as  $  (N/M)^{3/2}$, respectively \cite{NJP}, clearly preventing the production of $M\gg N$  high-fidelity clones.

\medskip 

{\em Probabilistic cloning of maximally entangled states and the optimality of universal gate replication.}    
The  map $  U_g  \to |\Phi_g\>$  is a one-to-one correspondence between qubit gates 
 and two-qubit maximally entangled states 
  \cite{choi,leung-thesis}.   The inverse map  $   |\Phi_g\>  \to  U_g$ is implemented by probabilistic teleportation,  which succeeds with optimal probability $1/4$   \cite{teleportation,chiribella-genkina-hardy}.  Combined  with the super-generation of entangled states, the  implementability of the map $   |\Phi_g\>  \to  U_g$  implies that  $N$ maximally entangled states can be cloned  probabilistically,  obtaining up to $N^2$ high fidelity copies,  albeit with  exponentially small probability $1/4^N$.  The result is interesting not only because it provides  an explicit protocol achieving super-replication of maximally entangled states, but also because it allows one to prove the optimality of the gate super-replication protocol:   if there existed a protocol producing $ M\gg N^{2}$ almost perfect copies of a generic unitary gate, such a protocol could be converted into a  probabilistic cloning protocol  producing $M\gg N^{2}$ almost perfect copies of a generic maximally entangled state. Such a protocol is impossible because it would violate the Heisenberg limit for quantum cloning \cite{stateREP}.       Even more strongly, since  the fidelity of state replication  must vanish for  $M\gg N^{2}$  \cite{stateREP},   every  gate replication protocol simulating  $M  \gg N^{2}$ uses   must have vanishing entanglement fidelity, and, therefore,  vanishing fidelity on most input states.  This conclusion    applies both  to deterministic and probabilistic gate replication protocols.

\medskip  

{\em Gate compression.}   The argument used to prove  gate super-replication can also  be applied to  the task of gate compression,  whose goal is  to encode the action of a gate   $U_x$  into another gate  $U_x'$ acting on a smaller physical  system.   
Gate compression protocols  are useful in   distributed scenarios wherein a server  (Alice)   is required  to apply the gate $U_x$   to an input state provided by a client (Bob).   In such a task, it is natural to minimize  the total amount of communication between client and server, by compressing the gate $U_x$   into a gate acting on the smallest possible system.        Ideally, the compression should be faithful, in the sense that the action of $U_x$ on the original input state is simulated without errors.  In practice, Alice and Bob can often tolerate a small error, especially if this allows them to increase the compression rate.     

\begin{figure}
  \centering
  \includegraphics[width=1.02\linewidth]{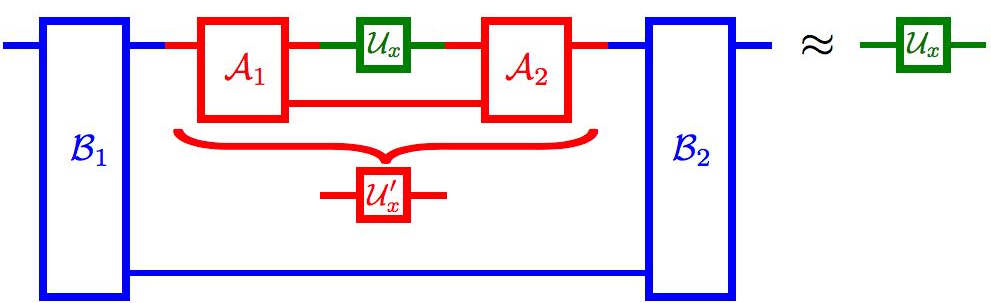}\\
  \caption{{\bf Gate compression.}  Alice sandwiches the given gate $\map U_x$ (in green) between two quantum channels $\map A_1$ and $\map A_2$  (in red), thus compressing it into a gate $\map U_x'$  acting on a smaller quantum system.  The action of $\map U_x$ is retrieved through  Bob's operations  $\map B_1$ and $\map B_2$ (in blue), which involve the system processed by Alice and a quantum memory  kept in Bob's laboratory.  The protocol  reduces the amount of quantum communication  needed to simulate the application of  Alice's gate to  Bob's input.      }  \label{fig:gatecomp}
  \end{figure}

  The general form of a gate compression protocol is illustrated in Figure \ref{fig:gatecomp}.  
  Here we consider the scenario where  $U_x$ is an $N$-qubit gate of the i.i.d. form $U_g^{\otimes N}$, $U_g$  being an arbitrary rotation of the Bloch sphere. The case of  rotations around a fixed axis has been previously considered in \cite{duer}. 
Using the decomposition of Eq. (\ref{spacedecomp}),  
the gate $U_g^{\otimes N}$ can be put in the block diagonal form  
$U_g^{\otimes N}   =  \bigoplus_{j=0}^{N/2}   \,    \left(     U_g^{(j)}  \otimes I_{\spc M_{jN}}\right)  $ ,
where $U^{(j)}_g$ is a unitary gate acting on the representation space $\spc R_j$ and $I_{\spc M_{jN}}$ is the identity on the multiplicity space ${\spc M_{jN}}$.  The working principle    of the  compression protocol is to get rid of the multiplicity spaces, on which the gate  $U_g^{\otimes N}$ acts trivially.  Specifically, Bob encodes his $N$-qubit input into the state of a composite system $AB$, where system $A$ has Hilbert space $\spc H_A    =   \bigoplus_{j  = 0}^{N/2}     \spc R_j$  and  system $B$ has Hilbert space $\spc H_B  = \spc M_{0N}$, the multiplicity space of largest dimension.
 Then, he  transmits system $A$ to Alice,  keeping system $B$ in his laboratory.  On her side, Alice encodes the gate   $U_g^{\otimes N}$ into the gate   $U_g'  =   \bigoplus_{j=0}^{N/2}     U_g^{(j)}$,  acting only on system $A$. She applies $U_g'$ on the input  provided by Bob and returns him the output.  Finally, Bob applies to systems $A$ and $B$  a joint decoding  operation, thus obtaining an output state of $N$ qubits.      All these operations   can be devised in such a way  that the protocol implements the gate $U_g^{\otimes N}$  \emph{exactly}  on every $N$-qubit input state   (Appendix D).    The key feature of this protocol is that Alice and Bob only communicate through the exchange of system $A$, whose dimension grows like $N^2$ instead of $2^N$. With respect to the naive protocol in which Alice and Bob send to each other the input and the output of the gate $U_g^{\otimes N}$, the protocol allows for an exponential reduction of quantum communication from  $2N$ qubits to  $4\log_2 N$ qubits.  
 The amount of quantum communication  can be further cut down by nearly a half if a small error is allowed.  Indeed, before sending the input to Alice, Bob can apply the encoding operation  $\map E^{(N)}_J$, which compresses the input into a subspace of dimension $O(J^2)$.     Setting    $J  =   \lfloor  \sqrt{N^{1+\delta}} \rfloor$ for some $\delta\in(0,1)$,  this encoding will cause little disturbance, except on an exponentially small fraction of the inputs (cf. theorem \ref{theo}).   As a result,  a completely unknown i.i.d.  sequence   of  $N$ qubit gates  on Alice's side can be approximately reproduced on Bob's side  through the exchange of   $2 (1+\delta)  \,   \log_2 N $ qubits, with  arbitrarily small $\delta>0$  (Appendix D).

{\em  Extension to higher dimensions.}   
Our gate replication protocol can be easily extended to quantum systems of arbitrary  dimension  $d<\infty$ (qudits), 
showing that arbitrary i.i.d. sequences  of unitary gates can be replicated at a nearly quadratic rate, with vanishing  error on most inputs except for an exponentially small fraction (Appendix E).  For    $M=  O(  N^{2-\alpha})$ the number of ancillary qudits used by the protocol is  equal to $M-N+   O\left(  N^{1-
\alpha/4} \right)$. As a byproduct of universal gate super-replication,  it is possible to achieve super-generation of maximally entangled states, as well as multiphase states of the form $|e_{\boldsymbol \theta}  \>  =       ( |0\>   +    e^{-i\theta_1}  \, |1\>  +   e^{-i\theta_2}  \,|2\>  +  \dots  +   e^{-i\theta_{d-1}}\,  |d-1\>)/\sqrt d$  (Appendix F).  
 Furthermore,  $N$ uses of a completely unknown gate can be compressed with zero error into a single gate acting on  $(d-1)(d/2+1)  \log_2 N$ qubits, a number that can be cut down  by nearly a half if one accepts an almost everywhere vanishing error  (Appendix G).
 
In conclusion, we showed that    $N$ uses of a completely unknown unitary gate allow one to simulate up to $N^2$ uses of the same gate, with high accuracy on all input states except for an exponentially small fraction. The protocol has optimal rate: any attempt to simulate   $N^2$  or more uses is doomed to have  vanishing fidelity on most  input states.    The ability to replicate unitary gates with high fidelity is not in contradiction with the  no-cloning theorem,   due to  the impossibility to deterministically retrieve unitary gates from non-orthogonal pure  states.  
The arguments developed for the gate replication protocol also  apply the task of gate compression, useful   in distributed scenarios where a server has to apply a gate to an input state provided by a client. In this scenario,   an unknown i.i.d. sequence of $N$ unitary gates can be compressed down to a single  gate acting only on $O(\log N)$ qubits, thus  achieving an exponential reduction of quantum communication    between client and  server.

\medskip  

{{\bf Acknowledgments.} 
 This work is supported by the National Basic Research Program of China (973) 2011CBA00300 (2011CBA00301),  by the National Natural Science Foundation of China through Grants  11450110096, 61033001,  and 61061130540,  by the Foundational Questions Institute through the Large Grant ``The fundamental principles of information dynamics",  and by the 1000 Youth Fellowship Program of China.  GC is grateful to P Hayden for pointing out the concentration result for the Schur-Weyl measure in Ref. \cite{harrowthesis} and to W D\"ur for valuable comments. }

\bibliography{gate}
\bibliographystyle{apsrev4-1}

\appendix

\begin{widetext}

\section{ Concentration of the Schur-Weyl measure for $d=2$}\label{app:fidelity}

Let us denote the  Schur-Weyl measure by $p_{j}  :  =  d_k  \, m_{jK}/d^K$. By definition,  one has
\begin{align*}
\nonumber p_{j}  &  =\frac{   d_j    \,  m_{jK}}{2^K}   \\
\nonumber & =  \frac{(2j+1)^2}{2^K(K/2+j+1)}   {K\choose K/2+j}  \\
\nonumber & \le      \, \frac {K+1} {2^K} {K\choose K/2+j}  
 \end{align*}
having used the expressions $d_j  =  2j+1$ and  
$m_{jK}=\frac{2j+1}{K/2+ j+1}{K\choose K/2+j}$. Then, the probability that the angular momentum number is no larger than  $J$ is lower bounded as  
\begin{align*}
\sum_{j=0}^J    p_j &\ge
  1  -     (K+1)  \,   \sum_{j>J}      \frac 1 {2^K} {K\choose K/2+j}  \\
  &  \ge    1 -       (K+1)   \,    \exp\left[   -  \frac{  2J^2}K \right]    \, ,
 \end{align*}
having used Hoeffding's inequality.  As a consequence, the entanglement fidelity of the channel $\map E^{(K)}_J$ is lower bounded as  
\begin{align}
\nonumber 
F^{(K,J)}_E   &     =  \left( \sum_{j=0}^{J}   p^{(K)}_{j}  \right)^2    \\
\label{FEbound} &\ge 1-          2(K+1) \,    \exp\left[   -  \frac{  2J^2}K \right]      \, .
\end{align}

\section{Minimum dimension of the ancilla in the gate replication protocol}
In the gate replication protocol, the output of the compression channel $\map E^{(M)}_J$ is encoded via an isometry     $V_J:   \spc H_J^{(M)}  \to  \spc H^{\otimes N}  \otimes \spc H_A$ satisfying 
 \begin{align}\label{comm1}  V_J  \, U_g^{\otimes  M}  =    \left(U_g^{\otimes N}  \otimes I_A\right) \, V_J   \qquad \forall\ U_g \in  \grp{SU}  (2) \, .
 \end{align}    
In order for such isometry to exist,  the  dimension of the ancilla must  satisfy the condition  
$m_{jN} \,  d_A  \ge   m_{jM}$ for all $j  \le J $. Hence, its  minimum value is 
\begin{eqnarray*}
d_A^{\min}&=    \left\lceil    \min_{j\le J}     \frac{m_{jM}}{m_{jN}}  \right\rceil \\
  &  =  \left\lceil   \frac{(N/2+J+1){M\choose M/2+J}}{(M/2+J+1){N\choose N/2+J}}\right\rceil \, .
\end{eqnarray*}  
For $J/N\ll 1  $, Stirling's approximation gives    $\log_2  d_A^{\min}   =    M-N  +  O(J^2/N)$.  The condition $J/ N\ll 1$  is always met in the super-replication regime, when $M$ scales like $N^{2-\alpha}$ for some $\alpha  >0$ and $J$ is set to $\left\lceil   N^{1-\alpha/4}\right\rceil$.  Accordingly, the number of ancillary qubits needed for the implementation  of the gate replication protocol is equal to 
\[  \log_2  d_A^{\min}  =    M-N    + O\left(   N^{1-\alpha/2}\right) \, .\]
\section{ Super-generation of two qubit maximally entangled states and single qubit phase states }  

Given $N$ uses of a generic qubit gate $U_g$, it is easy to obtain up to $N^2$ copies of the maximally entangled state $|\Phi_g\>  =    (U_g\otimes I )  |\Phi^+\>$, $|\Phi^+\> =   ( |0\>|0\>   +  |1\>  |1\>  )/\sqrt 2$, starting from     $N$ uses of the qubit gate $U_g$.   The protocol is as follows:  
\begin{enumerate}
\item use the gate replication protocol to simulate $M  $ uses of the gate $U_g$
\item apply the simulation of $U_g^{\otimes M}$ on the first qubit of the each pair in the entangled state $|\Phi^+\>^{\otimes M}$
\end{enumerate} 
The protocol produces an approximation of the target state $|\Phi_g\>^{\otimes M}$, given by
\begin{align*}
\rho^{(M)}_g   :  =   \left\{\left[   \map C_2    \,  \left(  \map U_g^{\otimes N}     \otimes \map I_A   \right)  \,  \map C_1   \right]    \otimes   \map I_{\rm sec} \right\} \,   \left(   |\Phi^+\>\<  \Phi^+|  \right)^{\otimes M}   \, ,
\end{align*}     
where   $  \map I_{\rm sec}$ denotes the  identity on the  $M$-qubit system consisting of the second qubit of each pair  in the entangled state $|\Phi^+\>^{\otimes M}$ and the channel $   \map C_2    \,  \left(  \map U_g^{\otimes N}     \otimes \map I_A   \right)  \,  \map C_1$ acts on the $M$-qubit system consisting of   the  first qubit of each pair (reordering of the Hilbert spaces is assumed where necessary).  
The fidelity of the approximation is   
\begin{align*}
F^{\rm ent}_{\rm gen}  [N\to M]  &=    \<   \Phi_g|^{\otimes M}   \,  \rho^{(M)}_g    \, |\Phi_g\>^{\otimes  M}\\
&  = \<   \Phi_g|^{\otimes M}   \left\{  \,\left[   \map C_2    \,  \left(  \map U_g^{\otimes N}     \otimes \map I_A   \right)  \,  \map C_1   \right]    \otimes   \map I_{\rm sec}   \right\} \,   \left(   |\Phi^+\>\<  \Phi^+|  \right)^{\otimes M}   \,      \, |\Phi_g\>^{\otimes  M}     \\
&=     \<   \Phi^+|^{\otimes M}    \,\left(   \map C_2    \,    \map C_1       \otimes   \map I_{\rm sec} \right) \,   \left(   |\Phi^+\>\<  \Phi^+|  \right)^{\otimes M}     \, |\Phi^+\>^{\otimes  M}     \\
&=     \<   \Phi^+|^{\otimes M}     \,\left(   \map E^{(M)}_{J}     \otimes   \map I_{\rm sec} \right) \,   \left(   |\Phi^+\>\<  \Phi^+|  \right)^{\otimes M}    \, |\Phi^+\>^{\otimes  M}  \\
&  \equiv    F_E^{(M,J)}   \\
\end{align*}
Setting $J  =  N/2$ and $K=M$, Eq. (\ref{FEbound}) then implies the bound 
\begin{align}\label{entangledstate} F^{\rm ent}_{\rm gen}  [N\to M]  \ge  1-  2  (M+1)\, \exp\left[-\frac{N^2}{2M}\right]    \, , 
\end{align}
which establishes the possibility to generate  up to $N^2$ copies of the state $|\Phi_g\>$ from $N$ uses of the gate $U_g$.  In other words, maximally entangled states can be super-generated using the corresponding gates.  

\medskip
Super-generation of entangled states also  implies super-generation of phase states.   Suppose that one is  given $N$ uses of a phase gate $  U_{\st n, \theta}$, where the rotation axis $\st n$ is known.   For definiteness, let us fix $\st n$ to be the $z$-axis, so that   $  U_{\st n, \theta}$ is diagonal in the computational basis $\{  |0\>, |1\>\}$.  Then, using the entanglement generation protocol, one can produce  $M$ approximate copies of the state $  |\Phi_\theta\>= ( |0\>|0\>   +  e^{-i \theta}|1\>  |1\>  )/\sqrt 2$.     The latter can be transformed into $M$ approximate copies of the state $|e_\theta\>  =( |0\>  +  e^{-i\theta}  |1\>  )/\sqrt 2$, by applying  ${\tt CNOT}$ gates  and discarding the second qubit of  each pair.  
Denoting by $  \rho^{(M)}_\theta$ the approximation of the $M$ entangled states,  the fidelity of the protocol is  given by
\begin{align}
\nonumber F^{\rm phase}_{\rm gen}  [N\to M] &  =  \<  e_\theta|^{\otimes M}    \Tr_{\rm sec }   \left[   {\tt CNOT}^{\otimes M}    \rho^{(M)}_\theta  {\tt CNOT}^{\otimes M}     \right ]  \,  |e_\theta\>^{\otimes M} \\
\nonumber  &  \ge    \left(\<  e_\theta|  \<  0|\right)^{\otimes M}    \,     {\tt CNOT}^{\otimes M}    \rho^{(M)}_\theta  {\tt CNOT}^{\otimes M}     \,     \left( |e_\theta\>   |0\> \right)^{\otimes M}    \\
\nonumber  & =   \<\Phi_\theta  |^{\otimes M} \,      \rho^{(M)}_\theta \,  |\Phi_\theta\>^{\otimes M}  \\
\nonumber  & \equiv F^{\rm ent}_{\rm gen} [N\to  M]  \\
\label{phasefid}&  \ge     1-  2  (M+1)   \,\exp\left[-\frac{N^2}{2M}\right]  \, .   
\end{align}
Note that the  protocol used to super-generate the phase state $|e_\theta\>$ is non-universal due to the presence of the $\tt CNOT $ gate, which is defined in the computational basis  $\{|0\>, |1\>\}$.

\section{Alice's and Bob's operations in the gate compression protocol}

Let us consider first the exact gate compression protocol, which faithfully encodes  the gate  $U_g^{\otimes N}$, acting on the Hilbert space
$ \spc H^{\otimes N}  =     \bigoplus_{j=0}^{N/2}    \left(    \spc R_j \otimes  {\spc M_{jN}} \right)  ,$
into the gate
$  U'_g  =     \bigoplus_{j=0}^{N/2}     U_g^{(j)}$, 
acting on the Hilbert space $\spc H_A =   \bigoplus_{j=0}^{N/2}   \spc R_j$.
In this case, Alice's operations are given by 
\begin{align}
\nonumber
\map A_1  (\cdot)     & :  =   V  \cdot V^\dag  \\
\label{aliceop}\map A_2  (\cdot)     &:  =    V^\dag  \cdot   V   +  \Tr[   (I  -  VV^\dag) \,    \cdot  ~ ]  \,  \alpha_0  \, ,         
\end{align}
where  $\alpha_0$ is a fixed density matrix on $\spc H_A$  and  $    V  :  \spc H_A \to \spc H^{\otimes N}  $ is the isometry 
$  V  : =   \bigoplus_{j=0}^{N/2}       \, I_{\spc R_j}  \otimes  |\mu_j\>  $,    where $ I_{\spc R_j} $ is the identity on $ {\spc R_j}  $ and $|\mu_j\>$ is a fixed pure state in $\spc M_{jN}$.    By construction, one has 
\begin{align}\label{enc}
\map A_2  ~  \map U_g^{\otimes N}   \map A_1     =   \map U_g'   \qquad \forall  U_g  \in  \grp{SU}  (2)  \, ,
\end{align}
where  $\map U_g$ and $\map U_g'$ are the  channels associated to  $U_g$ and $U_g'$, respectively. 
Eq. (\ref{enc}) expresses the fact that Alice's operations encode the action of the gate $U^{\otimes N}_g$ into the gate $U_g'$.   

In order to show that the encoding is faithful, one has to show that Bob can simulate  the gate $U^{\otimes N}_g$ using  $U_g'$. 
To this purpose, consider the operations  
\begin{align}
\nonumber
\map B_1  (\cdot)     & :  =   W  \cdot W^\dag  \\
\label{bobop}\map B_2  (\cdot)     &:  =    W^\dag  \cdot   W   +  \Tr[   (I  -  WW^\dag)  \, \cdot ~]  \,  \beta_0  \, ,         
\end{align}
where $\beta_0$ is a fixed density matrix on $\spc H^{\otimes N}$ and $W$ is the isometry 
\[    W:  \spc H^{\otimes N}  \to   \spc H_A  \otimes \spc H_B  \, ,   \qquad W : = \bigoplus_{j=0}^{{N/2}}       \left(   I_{\spc R_j}  \otimes      W_j    \right)    \,, \]
where $\spc H_B  :  =  \spc M_{0N}$ is the multiplicity space of largest dimension and  $W_j$ is a fixed isometry from  $\spc M_{jN}$ to $\spc H_B$. With this definition, one has 
 \begin{align}\label{dec}
\map B_2  ~  \left(\map U_g'   \otimes \map I_B   \right)  \,\map B_1     =   \map U_g^{\otimes N}   \qquad \forall  U_g  \in  \grp{SU}  (2)  \, ,
\end{align}
expressing the fact that Bob can retrieve the action of the gate $U_g^{\otimes N} $ from the use of $U_g'$. According to Eqs. (\ref{enc}) and (\ref{dec}), the operations $\map A_1,\map A_2,\map B_1,\map B_2$ define an exact gate compression protocol. 

The approximate protocol is a straightforward variation of the above: the gate $U_g^{\otimes N}$ is encoded into the gate $U_g'  =  \sum_{j=0}^J   U_g^{(j)}$, with  $J  =     \lfloor   \sqrt{N^{1+\delta}} \rfloor $ for some  $\delta \in  (0,1)$. Alice's operations are of the same form as  those in Eq. (\ref{aliceop}), with the only difference that now the domain of the isometry $V$  is not the whole space  $ \spc H_A$, but rather the subspace $\spc H_{A,J}  := \sum_{j=0}^J   \spc R_j $. 
   Bob's operations are of the same form as those in Eq. (\ref{bobop}), with the difference that the operation $\map B_1$ is replaced by $\map B_{1,J}  :  =    \map B_1   \,\map  E_J^{(N)}$, which outputs states in $\spc H_{A,J}  \otimes \spc H_B$.    
  
\section{Universal gate replication in dimension $d\ge 2$}

The gate replication protocol for qudits   is the immediate generalization of the protocol for qubits. The main steps in the construction of the protocol are as follows:  First,  we define a suitable subspace   $\spc H^{(M)}_J  \subseteq  \spc H^{\otimes M}$ and construct the encoding operation  
\begin{align}\label{encodingd} \map E^{(M)}_J(  \cdot )  =    P^{(M)}_J   \cdot  P^{(M)}_J     +    \Tr  \left[ \left(  I^{\otimes M}   -  P^{(M)}_J   \right)  \cdot \right]  \,  \rho_0    \, ,
\end{align} 
where   $P_J^{(M)}$ is the projector on $\spc H_J^{(M)}$ and $\rho_0$ is some fixed state with support in $\spc H_J^{(M)}$.  
 We then show that, provided that $J$ is large enough,  the encoding channel induces little disturbance on most input states, with the  exception of an exponentially small fraction.   
 Furthermore, we show that the output of the encoding channel can be faithfully encoded into a composite system consisting of $N$ identical qudits and an ancilla $A$,  via an isometry $V_J  :  \spc H^{(M)}_J  \to    \spc H^{\otimes N}  \otimes \spc H_A $ satisfying the condition  
\begin{align}\label{commu}      V_J     U_g^{\otimes M}   =   \left(    U_g^{\otimes N}  \otimes I_A\right)    V_J   \, ,    \qquad  \forall  U_g  \in  \grp {SU} (d)  \, .  
\end{align}   
The action of the isometry $V_J$ can be inverted  by the decoding channel   
\begin{align}\label{decodingd}
  \map D_J(\rho)   :=   V_{J}^\dag  \rho  V_{J}    +      \Tr\left[  \left( I^{\otimes N}  -   V_{J}  V_{J}^\dag  \right)  \rho  \right]  \,  \rho_0 \,.
  \end{align}   
Given these ingredients, the gate replication protocol follows  the same steps as the gate replication protocol for qubits:
\begin{enumerate}
\item Send the $M$ input systems to the input of the  encoding channel $\map E^{(M)}_J$ with   $J   =     \left\lceil  N^{1-\alpha/4}\right\rceil$ for $M$ growing as $N^{2-\alpha}$, $\alpha  >0$ and $J=  N/d$ for $M$ growing as $N^{2}$ or faster 
\item  Apply the isometry $V_J$ to  the output of   $\map E^{(M)}_J$, thus encoding it  into the Hilbert space  $\spc H^{\otimes N}  \otimes \spc H_A$ 
\item Apply the gate $U^{\otimes N}_g$  on the   $N$ systems
\item Send the $N$ systems and the ancilla to the input of  the decoding channel $\map D_J$. 
\end{enumerate}

In the following we show how to construct the encoding map $\map E^{(M)}_J$ and the isometry $V_J$,  proving that the above protocol can simulate  $ N^2$ parallel uses of the gate  $U_g$ with high fidelity on most inputs.

   \subsection{Decomposition of the Hilbert space}  
 
Let $U_g$ be a generic element of the group $\grp {SU}(d)$, parametrized by a suitable vector $g  \in  \mathbb R^{d^2-1}$. 
 For a given integer $K\ge 0$, the irreducible representations in the decomposition of $U_g^{\otimes K}$ are  labelled by Young diagrams   with $K$ boxes arranged into $d$ rows.   A Young diagram is completely specified by the lengths of its rows, which can be put into a vector $\lambda  =  (\lambda_1,\dots,\lambda_d)$ of non-negative integers satisfying the conditions
 \begin{align}
 \lambda_1  \ge  \lambda_2  \ge  \dots \ge \lambda_d  \ge 0 \, ,  \qquad  
 \sum_{i=1}^d   \lambda_ i       =   K  \, . 
\end{align}
We denote by $\mathsf Y_{K,d}$ the set of all such vectors and,   from now on,  we  identify Young diagrams with the corresponding vectors, referring to $\lambda$ as a ``Young diagram".     
 Note that one has  \begin{align}\mathsf Y_{K,d} \subset \mathsf{T}_{K,d} \, ,
\end{align}
where $\mathsf{T}_{K,d}$  is the set of all partitions of $K$ into $d$ non-negative integers.  

\medskip 
With the above notation,   the Hilbert space of $K$ identical systems can be decomposed as
\begin{align}\label{spacedecomp1}
\spc H^{\otimes K}   \simeq \bigoplus_{\lambda\in\mathsf{Y}_{K,d}}    \,  \left(    \spc R_\lambda  \otimes \spc M_\lambda   \right)  \, ,
\end{align}
where $\spc R_\lambda $ is a representation space, of dimension $d_\lambda $, and  $\spc M_\lambda$ is the corresponding multiplicity space, of dimension $m_\lambda$.  The dimensions and multiplicities satisfy the following bounds \cite{hay,CM,harrowthesis}.   
\begin{align}\label{dilambda}
d_\lambda  \le (K+1)^{\frac{d(d-1)}{2}}
\end{align}
and
\begin{align}\label{mult}
 {K\choose \lambda}(K+1)^{-\frac{d(d-1)}{2}}\le  ~ m_\lambda  ~\le {K\choose \lambda}  \, , 
\end{align}
where $ {K  \choose  \lambda  } :  =   \frac{K!}{\lambda_1!  \dots  \lambda_d!} $  is the multinomial coefficient.  

\medskip

 Relative to the decomposition (\ref{spacedecomp1}), the gate $U_g^{\otimes K}$  can be written  in the block diagonal form
\begin{align*}
U_g^{\otimes K}   \simeq \bigoplus_{\lambda\in\mathsf{Y}_{K,d}}    \,  \left[       U^{(\lambda)}_g    \otimes I_{\spc M_\lambda}   \right]  \, ,
\end{align*}  
where $U_g^{(\lambda)}    $ is an irreducible representation (irrep) of $\grp {SU}  (d)$ and $ I_{ \spc M_\lambda}   $ is the identity matrix on  $\spc M_\lambda$.   

Different Young diagrams in $\mathsf Y_{K,d}$ correspond to different irreps.   However, for $K  \not  =  K'$ two  Young diagrams  $\lambda \in \mathsf Y_{K,d}$ and $\lambda'  \in  \map Y_{K',d}$  can correspond to the same irrep, provided that one has 
\begin{align}\label{equivalent}
\lambda   =       \lambda'   +   \lambda_0 \, {\bf 1} \, ,     
\end{align}
where $\lambda_0   \in  \mathbb Z$ is some fixed integer and ${\bf 1} \in  \mathbb R^d$ is the vector with all entries equal to $1$.

\subsection{Concentration of the Schur-Weyl measure}

The Schur-Weyl measure is the probability distribution over the Young diagrams in $\mathsf Y_{K,d}$ defined as      
 \[p_\lambda:  =  \frac{d_\lambda\, m_\lambda}{d^K}  \,  .\]  
 For large $K$, the Schur-Weyl measure  is concentrated on the Young diagrams with rows of length approximately equal to $K/d$   \cite{harrowthesis}.  Specifically, let us denote by $\mathsf{Y}^{(K)}_{J}$  the set of  Young diagrams with the last row no shorter  than $K/d-  J$,~i.~e.
\begin{align}\label{youngest}
 \mathsf{Y}^{(K)}_{J}  :  =   \left  \{     \lambda\in  \mathsf Y_{K,d}   ~\left|~   \lambda_d  \ge \frac K d    -  J  \right  \} \right. 
 \end{align}
and by $\overline {\mathsf Y}^{(K)}_{J}  :  =   \map  Y_{K,d}  \setminus  \mathsf{Y}^{(K)}_{J} $ its complement.  Then, we have  the following  
\begin{lem}
The Schur-Weyl measure of $\overline {\mathsf Y}^{(K)}_{J}  $ is upper bounded as 
\begin{align}\label{conce}
{\sf Prob}  \left[  \lambda \in  \overline {\mathsf Y}^{(K)}_{J}    \right]  \le    (K+1)^{\frac{d(d-1)}{2}}   \, \exp  \left[   - \frac{ 2  J^2} K \right]      \, .
\end{align}
\end{lem}
{\bf Proof.}     Using Eqs.  (\ref{dilambda}) and (\ref{mult}),  we obtain the bound
\begin{equation}\label{multi}
  {\sf Prob}  \left[  \lambda \in  \overline {\mathsf Y}^{(K)}_{J}    \right]    \le
       (K+1)^{\frac{d(d-1)}{2}}     ~   
\sum_{\lambda\in\overline{\mathcal{Y}}^{(K)}_J}  q_\lambda    \, , 
\end{equation}
where $q_\lambda:  =   \frac1{  d^K  }   {K  \choose \lambda}$ is the multinomial distribution.   In turn, the summation in the r.h.s. of Eq. (\ref{multi}) can be upper bounded by extending the range    from  Young diagrams   to general partitions of $K$,  as follows
\begin{align*}
\sum_{\lambda\in\overline{\mathcal{Y}}^{(K)}_J}       q_\lambda   
&\le    
\sum_{
\begin{array}{l}  
\lambda\in\mathsf T_{K,d}  \, ,  \\
\lambda_d  <  \frac Kd  - J
\end{array}}
   \, q_\lambda  \\
  &   = \sum_{\lambda_d   < \frac Kd  -  J}   \,  \left(\frac 1 d\right)^{\lambda_d}   \,   \left(   1-  \frac 1d\right)^{K-\lambda_d}\,  \begin{pmatrix}   K  \\ \lambda_d\end{pmatrix}   \\
 &\le    \exp  \left[   - \frac{ 2  J^2} K \right]    \, ,
\end{align*}
having used Hoeffding's inequality.  Combining the above bound with Eq. (\ref{multi}) one obtains the desired result. 
 \qed

 \subsection{The encoding operation}
  Define the subspace   $\spc H^{(K)}_J  \subseteq  \spc H^{\otimes K}$ as  
\begin{align}\label{spcJ} \spc H^{(K)}_J  :  =   \bigoplus_{\lambda   \in    \mathsf{Y}^{(K)}_{J}}     \left(\spc{R}_\lambda  \otimes\spc{M}_\lambda\right)       
\end{align}
and the encoding operation $\map E^{(K)}_J$  as in Eq. (\ref{encodingd}).  
 Denote  by $F^{(K,J)}_E$  be the  entanglement fidelity of $\map E^{(K)}_J$,    given by  \[  F^{(K,J)}_E  =   \<  \Phi_{d^K} | \left(   \map E^{(K)}_J  \otimes \map I^{\otimes K}  \right)   (|\Phi_{d^{K}}\>\<\Phi_{d^K}|) \,   |\Phi_{d^K}\>   \, ,\]   $|\Phi_{d^K} \>$ being  a maximally entangled state in $\mathbb C^{d^K}\otimes \mathbb C^{d^K}$.
Then, we have the following 
\begin{lem}\label{lem:concen} 
The entanglement fidelity  of channel $\map E^{(K)}_J$ is lower bounded as
\begin{align}\label{entfid1}
F_E^{(K,J)}&\ge    1  -   2   \,          (K+1)^{\frac{d(d-1)}{2}}   \,     \exp  \left[   - \frac{ 2  J^2} K \right] \,.
\end{align}
\end{lem}

{\bf Proof.}    By definition of the encoding channel, the entanglement fidelity satisfies  the bound  
\begin{align}
\nonumber F^{(K,J)}_E  &\ge   \left |  \<   \Phi_{d^K}  |    \left(  P_J^{(K)} \otimes I^{\otimes K}\right)\,  |\Phi_{d^K}\>  \right|^2 \\
\nonumber & = \left[ \sum_{\lambda \in\mathsf{Y}^{(K)}_{J}}   \frac{d_\lambda m_\lambda}{2^K}\right]^2 \\
\nonumber  & \equiv   \left\{    {\sf Prob}    \left[  \mathsf{Y}^{(K)}_{J}    \right]  \right\}^2\\
\nonumber  &\ge   1  -   2       {\sf Prob}    \left[  \overline{\mathsf Y}^{(K)}_{J}    \right]     \, . 
\end{align}
Inserting Eq.  (\ref{conce}) in the bound one obtains the desired result. \qed

\medskip

Using the bound on the entanglement fidelity, it is immediate to obtain a bound on the probability that  
  a random   $K$-partite   state  $|\Psi\> $  has high fidelity with the state $  \map E^{(K)}_J  (  |\Psi\>\<\Psi|)$.   
  The result is a generalization of theorem 1 in the main text, which now reads 
\begin{theo}\label{theo1}   
Let  $  F_\Psi^{(K,J)}$ be the fidelity between the pure state $|\Psi\>\in\spc H^{\otimes K} $ and the state $  \map E^{(K)}_J  (  |\Psi\>\<\Psi|)$.   
If $|\Psi\>$ is chosen uniformly at random, then one has
\begin{align}\label{smallprob1}
{\sf Prob}   \left[ F^{(K,J)}_\Psi   <  1-\epsilon \right]  <   
2   \,          (K+1)^{\frac{d(d-1)}{2}}   \,       \,    \frac{         \exp\left[-\frac{2J^2}{K}   \right]}{\epsilon}   
\end{align}
 for every fixed $\epsilon>0$. 
 \end{theo}

{\bf Proof.}       Recall that the average of fidelity over all pure states can be is lower bounded by the entanglement fidelity.   Combining this fact  with Markov's inequality
 one obtains 
\[   {\sf Prob}   \left[ F^{(K,J)}_\Psi   <  1-\epsilon \right]    <      \,   \frac{1-      F^{(K,J)}_E  }  \epsilon   \, .   \] 
Inserting Eq. (\ref{entfid1})  in the above bound one obtains the desired result.  \qed 

\medskip  

The construction of the gate replication protocol uses theorem \ref{theo1} with $K  = M$.       The theorem will be  used also in the approximate gate compression protocol, in that case by setting  $K=  N$.

\subsection{Embedding into the space of $N$ systems}

We now construct the isometry $V_J $, which embeds the subspace $\spc H_J^{(M)}$ into the Hilbert space of $N$ identical copies and a suitable ancilla.  Here we assume that $M- N$ is a multiple of $d$.   Let us decompose the target Hilbert space as 
\begin{align}
\spc H^{\otimes N}  \otimes \spc H_A    \simeq    \bigoplus_{\lambda \in  \map Y_{N,d}}  \left(   \spc R_\lambda \otimes  \spc M_\lambda  \otimes \spc H_A  \right)\, .         
\end{align} 
By Eq. (\ref{commu}),   the isometry $V_J$ must be of the form  
\begin{align}\label{form}
V_J     :   =  \bigoplus_{\lambda\in  \map Y^{(M)}_{J}}  \,     \left( I_{\lambda}  \otimes   V_{\lambda} \right)\, ,
\end{align} 
 where   $  I_{\lambda}$ is a unitary isomorphism between  the representation space $  {\spc R_\lambda}$ and the representation space  $\spc R_{\lambda'}$, with
 \[  \lambda'   :  =     \lambda   -    \frac{ M-N}d  \,  {\bf 1} \, ,\]
 and $V_\lambda$ is an isometry from   $  \spc M_\lambda  $ to $  \spc M_{\lambda'} \otimes   \spc H_A$.  
 In order for Eq. (\ref{form}) to hold, two conditions must be met:    
 \begin{enumerate}
 \item all the irreducible representations corresponding to Young diagrams in $\map Y^{(M)}_J$ should be contained in the decomposition of $U_g^{\otimes N}$
 \item  the dimension of   $\spc M_{\lambda'} \otimes   \spc H_A$ should be larger than the dimension of $  \spc M_\lambda  $ for every $\lambda  \in\map Y^{(M)}_J$.  
 \end{enumerate} 
 Condition 1 is equivalent to the requirement that every Young diagram   $\lambda \in \map Y^{(M)}_{J}$ be of the form
 \begin{align}
\lambda  =   \frac{ M-N} d \,  {\bf 1}  +   \lambda'   \qquad \lambda'  \in  \map Y^{(N)}_J \, . 
\end{align}
The minimum ancilla dimension compatible with Condition 2   is  then given by  
  \begin{align}\label{damax}
  d^{\min}_{A}      &   =   \max_{\lambda' \in  \map Y^{(N)}_J}   \,  \left \lceil  \frac{ m^{(M,d)}_{\frac{ M-N}d   \,  {\bf 1}  + \lambda'} }{m^{(N,d)}_{\lambda'}    } \right\rceil \, .
  \end{align}
Using Eq. (\ref{damax}) we can estimate how many ancillary  qudits are required.    In the  super-replication regime (i.~e.~when $M$ grows as $N^{2-\alpha}$ for some $0<\alpha<1$),  the minimum number  is asymptotically equal to $M-N$, up to terms that are negligible compared to $N$.   Indeed, we can use the bound 
\begin{align}\label{ancibound}
   (M+1)^{-\frac{d(d-1)}{2}}\,      \max_{\lambda' \in  \map Y^{(N)}_J}   \frac{{M\choose {  \frac{ M-N} d  \,  {\bf 1 }+  \lambda'}}}{{N\choose \lambda'} }  \le  d_A^{\min}    \le     (N+1)^{\frac{d(d-1)}{2}}\,    \max_{\lambda' \in  \map Y^{(N)}_J}   \frac{{M\choose {  \frac{ M-N} d  \,  {\bf 1 }+  \lambda'}}}{{N\choose \lambda'} }   \, ,   
 \end{align}
   following from Eq.(\ref{mult}).   Setting  $J  =   \left  \lceil N^{1-\alpha/4}  \right\rceil$, we can use  Stirling's approximation, which   for $J/K\ll 1$, $K  = M,N$,     yields
\begin{align}
\log_d {  K  \choose  \lambda}   =  K  -O(J^2/K)  
 \, ,  \qquad \forall  \lambda \in   \mathsf{Y}^{(K)}_{J}  \, .
 \end{align}
Inserting the above approximation in  the bounds of Eq. (\ref{ancibound}), we finally obtain the equality
$\log_d d^{\min}_A  =   M-N   +    O(J^2/N)$. 
which, recalling that $J$ was set to   $ \left  \lceil N^{1-\alpha/4}  \right\rceil$, yields
\begin{align*}
\log_d d^{\min}_A  =   M-N   +    O\left(N^{1-\alpha/2}\right) \, .
\end{align*}
Te above expression quantifies the number of ancillary qudits needed to achieve super-replication.  
\subsection{The fidelity of gate super-replication}
 We are now ready to evaluate the fidelity of the universal gate replication network.   Let us set    $\map C_1   :  =     \map V_J  \, \map E^{(M)}_J    $ and   $
\map C_2   :  =   \map D_J  $, where $\map E^{(M)}_J$ is the encoding channel of Eq. (\ref{encodingd}) and  $\map D_J$ is the decoding map of Eq. (\ref{decodingd}), which inverts the isometric channel $\map V_J  $.      Applying the gate replication protocol to a generic pure state $  |\Psi\>  \in  \spc H^{\otimes M}$, one obtains the output state
\begin{align*}
  \map C_2   (\map U_g^{\otimes N}  \otimes \map I_A)  \map C_1   (|\Psi\>\< \Psi| )  &  =    \map U_g^{\otimes M}   \map C_2  \map C_1  (  |\Psi\>\<  \Psi|)  \\
 & =   \map U_g^{\otimes M}     \map E_{J}^{(M)} (   |\Psi\>\<  \Psi|  )   
  \end{align*}
  for every gate $  U_g\in\grp {SU}(d)$. 
   Like in the qubit case,    the fidelity between the output state  and the ideal target $U_g^{\otimes M}  |\Psi\>$  is equal to the fidelity between $  \map E_{J}^{(M)} (  |\Psi\>\<  \Psi|)$ and $|\Psi\>$, equal to $F_\Psi^{(M,J)}$.   For $M   =  O(  N^{2-\alpha})$, $\alpha  >0$,  the choice $J    =   \left \lceil  N^{1-\alpha/4 } \right \rceil $, guarantees that the fidelity is arbitrarily close to 1 on all states except a low probability  subset:  
  for every fixed $\epsilon  > 0$, one has
 \begin{align}
{\sf Prob}   \left[ F^{(M,J)}_\Psi   <  1-\epsilon \right]  <    2   \,         (M+1)^{\frac{d(d-1)}2} \,   
  \frac{\exp \left[ -2   \sqrt {N^{\alpha}} \right] }{\epsilon}     \, ,
\end{align}
having used Eq. (\ref{smallprob1}). 
 
\section{Super-generation of maximally entangled states and multiphase states}  

Like in the qubit case, gate replication can be used to generate up to $N^2$ copies of a generic maximally entangled state starting from $N$ uses of the corresponding gate.  Setting $J  =    N/d$  in the gate replication protocol and following the same steps that led to the derivation of Eq. (\ref{entangledstate}) we obtain  
\begin{align*}
F^{\rm ent}_{\rm gen}[N\to M] &   =   F^{(M,N/d)}_E  \\
&  \ge      1  -   2    \,       (M+1)^{\frac{d(d-1)}{2}}   \,       \exp  \left[   - \frac{ 2  N^2} {d^2  M} \right]      \, .    
\end{align*}
In addition, the protocol can be easily adapted in order to achieve super-generation of a generic multiphase state  
\[   |e_{\boldsymbol \theta}  \>    =   \frac{   |0\>   +    e^{-i\theta_1}  \, |1\>  +   e^{-i\theta_2}  \,|2\>  +  \dots  +   e^{-i\theta_{d-1}}\,  |d-1\>}{\sqrt d}    \qquad   {\boldsymbol \theta} \in  [0,2\pi)^{\times (d-1)}  \, ,\]
starting from $N$ uses of the multiphase gate
$   U_{\boldsymbol \theta}   =   |0\>\<0|   +    e^{-i\theta_1}  \, |1\>\<1|  +   e^{-i\theta_2} \,  |2\>\<2|  +  \dots  +   e^{-i\theta_{d-1}}  \,|d-1\>\<d-1|$.  
 Indeed, it is enough to 
 \begin{enumerate}
 \item generate  $M$ copies of the entangled state  $ |\Phi_{\boldsymbol \theta}  \>    =   (  |0\> |0\>  +    e^{-i\theta_1}  \, |1\>|1\>  +   e^{-i\theta_2}  \,|2\>|2\>  +  \dots  +   e^{-i\theta_{d-1}}\,  |d-1\>|d-1\>)/\sqrt d $, 
 \item apply the inverse of the control-shift gate ${\tt C-SHIFT} $  on each pair of qubits, where ${\tt C-SHIFT}  := \sum_{n=0}^{d-1}    |n\>\<n|  \otimes    S^n$,  $S$ being is the cyclic shift on the computational basis, and 
 \item discard the second qubit of each pair.  
 \end{enumerate} 
 The same steps that led to Eq. (\ref{phasefid}) show that the fidelity of the above protocol is lower bounded as \[  F^{\rm multiphase}_{\rm gen}  [N\to M]    \ge 1  -   2 \,        (M+1)^{\frac{d(d-1)}{2}}   \,\,       \exp  \left[   - \frac{ 2  N^2} {d^2  M}  \right]  \, ,  \]
 thus guaranteeing super-generation of multiphase states.    
\section{Universal gate compression in dimension $d  \ge 2$}\label{app:gatecomphigh}  
The  gate compression protocols for qudits are of the same form of the protocols for qubits, with the only difference that the angular momentum number $j $  is replaced by the  vector $\lambda$ that parametrizes the Young diagrams. Here we quantify  the compression rates achieved for qudits.  

Let us start from  zero-error compression.      
The  protocol encodes the gate 
 \begin{align*}
U_g^{\otimes N}   \simeq \bigoplus_{\lambda\in\mathcal{Y}_{N,d}}    \,  \left[       U^{(\lambda)}_g    \otimes I_{\spc M_\lambda}   \right]  \, ,
\end{align*}   
into the  gate      $ U_g'    :  =  \bigoplus_{\lambda \in   \map Y_{N,d}}        U_g^{(\lambda)} $, 
 acting  on the smaller Hilbert space 
$\spc H_A  =  \bigoplus_{\lambda\in  \mathcal{Y}_{N,d}}\spc{R}_{\lambda}$.    
The dimension of $\spc H_A$ can be upper bounded as  
\begin{align}
\nonumber d_A   &=   \sum_{\lambda  \in  \map Y_{N,d}}     \,  d_\lambda  \\
 \nonumber   &  <  \,   \left(  \max_{\lambda \in \map Y_{N,d}}  d_\lambda\right)  \,      \left|   \map Y_{N,d}    \right|  \\
 \label{da}  &  \le   (N+1)^{\frac{d(d-1)}{2}}   \,      \left|   \map Y_{N,d}    \right|     \, ,    
\end{align}
having used Eq. (\ref{dilambda}) in the last inequality.  
 Since the total number of Young diagrams is smaller than the number of partitions in $\map T_{N,d}$,
 one has 
\begin{align*}  
|\mathcal{Y}_{N,d}|  &<  \begin{pmatrix}     N +  d-1 \\  d-1\end{pmatrix} \\
&    < (N+1)^{d-1}   \, ,
  \end{align*}
which, inserted in Eq. (\ref{da}), yields 
\begin{align}\label{compressed}
d_A  <  (N+1)^{(d-1)(d/2+1)}    \, .  
\end{align}  
Hence,  $N$ uses of a generic qudit gate can be compressed without errors into a single gate acting on $ (d-1)(d/2+1)  \log_2 N$ qubits in the large $N$ limit. 

Like in the qubit case, the approximate protocol is a straightforward variation of the zero-error protocol.  In the approximate version,  the gate $U_g^{\otimes N}$ is encoded into the gate $U_g'  =  \bigoplus_{\lambda  \in  \map Y^{(N)}_J}   U_g^{(\lambda)}$, with $J  =      \lfloor \sqrt{N^{1+\delta}} \rfloor$ for some $\delta  \in (0,1)$.   Now, all the elements of $\map Y^{(N)}_J$ are of the form  
\[\lambda  =      \frac{  N-N'}d\,  {\bf 1}   +  \lambda'   \qquad N'  :=  dJ  \, , \quad \lambda'  \in  \map Y_{N',d} \, .  \] 
Replacing $N$ with $N'$ in the steps leading to Eq. (\ref{compressed}) we obtain  $d_A  < \left [d   ( \sqrt{  N^{1+\delta}} +1)\right]^{(d-1)(d/2+1)} $, meaning that the gate $U_g^{\otimes N}$ can be compressed to a gate acting on  $(d-1)(d/2+1)(1+\delta)/2  \log_2 N$ qubits in the asymptotic limit.    

\end{widetext}
\end{document}